\documentclass[twocolumn]{aastex63}

\newcommand{\oii}{[\ion{O}{2}]}
\newcommand{\oiii}{[\ion{O}{3}]}
\newcommand{\sii}{[\ion{S}{2}]}
\newcommand{\siii}{[\ion{S}{3}]}
\newcommand{\ciii}{\ion{C}{3}]}
\newcommand{\nii}{[\ion{N}{2}]}
\newcommand{\hii}{\ion{H}{2}}
\newcommand{\Halpha}{{\rm H$\alpha$}}
\newcommand{\Hbeta}{H$\beta$}
\newcommand{\wavelength}[1]{$\lambda#1$}

\shorttitle{Limits to Ionization-Parameter Mapping}
\shortauthors{Sawant et al.}

\received{24 May 2021}
\revised{13 Aug 2021}
\accepted{\today}

\graphicspath{{./}{figures/}}

\begin{document}

\title{Limits to Ionization-Parameter Mapping as a Diagnostic of \hii\ Region Optical Depth}

\correspondingauthor{Amit N. Sawant}
\email{ansawant@umich.edu}

\author{Amit N. Sawant}
\affiliation{Department of Statistics, University of Michigan, 1085 South University Ave.,
Ann Arbor, MI   48109-1107, USA}
\affiliation{Department of Astronomy, University of Michigan,
1085 South University Ave.,
Ann Arbor, MI   48109-1107, USA}

\author[0000-0003-3119-0815]{Eric W. Pellegrini}
\affiliation{Zentrum für Astronomie, Institut für Theoretische Astrophysik, Universität Heidelberg,
Albert-Ueberle-Str. 2, D-69120 Heidelberg, Germany}

\author[0000-0002-5808-1320]{M. S. Oey}
\affiliation{Department of Astronomy, University of Michigan,
1085 South University Ave.,
Ann Arbor, MI   48109-1107, USA}

\author{Jes\'us L\'opez-Hern\'andez}
\affiliation{Facultad de Ciencias de la Tierra y el Espacio, Universidad Aut\'onoma de Sinaloa,
Av. Universitarios S/N, C. Universitaria, 80010, Culiac\'an, Sinaloa, M\'exico}

\author[0000-0003-4376-2841]{Genoveva Micheva}
\affiliation{Leibniz-Institute for Astrophysics Potsdam, An der Sternwarte 16, D-14482 Potsdam, Germany}

\begin{abstract}

We employ ionization-parameter mapping (IPM) to infer the optical depth of \hii\space regions in the northern half of M33.  We construct \oiii\wavelength{5007}/\oii\wavelength{3727}\space and \oiii\wavelength{5007}/\sii\wavelength{6724}\space ratio maps from narrow-band images continuum-subtracted in this way, from which we classify the \hii\space regions by optical depth to ionizing radiation, based on their ionization structure. This method works relatively well in the low metallicity regime, $12 + \log(\rm O/H) \leq 8.4$, where \oiii$\lambda\lambda4949,5007$ is strong.  However, at higher metallicities, the method breaks down due to the strong dependence of the \oiii$\lambda\lambda4959,5007$ emission lines on the nebular temperature.  Thus, although O$^{++}$ may be present in metal-rich \hii\ regions, these commonly used emission lines do not serve as a useful indicator of its presence, and hence, the O ionization state.  In addition, IPM as a diagnostic of optical depth is limited by spatial resolution. We also report a region of highly excited \oiii\space extending over an area $\sim$ 1 kpc across and \oiii$\lambda5007$ luminosity of $4.9\pm 1.5\times10^{38}$ erg/s, which is several times higher than the ionizing budget of any potential sources in this portion of the galaxy.  Finally, this work introduces a new method for continuum subtraction of narrow-band images based on the dispersion of pixels around the mode of the diffuse-light flux distribution.  In addition to M33, we demonstrate the method on \ciii$\lambda$1909 imaging of Haro~11, ESO 338-IG004, and Mrk~71.

\end{abstract}

\section{Introduction} \label{sec:intro}

The nebular ionization parameter describes the ionizing photon density relative to gas density, and it is a fundamental diagnostic of radiation feedback in photoionized \hii\ regions.  In recent years, diagnostics of the ionization parameter such as \oiii$\lambda5007$/\oii$\lambda3727$, \siii$\lambda9069$/\sii$\lambda6717,6731$, \oiii/H$\beta$, and \oiii/\sii\ \citep{Pellegrini2012ApJ...755...40P, Zastrow2013ApJ...779...76Z,Keenan2017ApJ...848...12K, Wang2019ApJ...885...57W} have been used to evaluate the nebular optical depth to Lyman continuum (LyC) radiation in both individual \hii\ regions and starburst galaxies.  An especially compelling class of objects are the Green Pea galaxies \citep{Cardamone2009MNRAS.399.1191C}, which are selected on the basis of their extreme ionization parameters in \oiii/H$\beta$.  Confirming predictions \citep[e.g.,][]{JaskotOey2013ApJ...766...91J}, the Green Peas have yielded the most consistent detections of LyC-emitting galaxies in the local universe \citep[e.g.,][]{Izotov2016MNRAS.461.3683I, Izotov2018MNRAS.478.4851I}, and are therefore of vital interest to galaxy evolution and cosmic reionization.  However, although we noted above a direct link between ionization parameter and LyC optical depth, the exact relationship between these  these quantities is not well understood in these starbursts, due to complicating factors like gas morphology, composition, geometry, and density distributions; and also variations in ionizing spectral energy distributions (SEDs) from various candidate stellar populations and other ionizing sources.  Nevertheless, the ionization parameter is an easily observed and widely used diagnostic of the nebular conditions in star-forming regions both near and far.

It is well known that \oiii$\lambda\lambda4959,5007$ emission drops precipitously at oxygen abundances $12+\log\rm(O/H) > 8.4$, \citep[e.g.,][]{Kewley2002ApJS..142...35K}. For example, the well-known abundance diagnostic $R23\equiv($\oii$\lambda3727 + $\oiii$\lambda\lambda4959,5007)/\rm H\beta$ increases monotonically to maximum values around this metallicity.  This is caused by strong sensitivity of this line to the electron temperature, which decreases at higher oxygen abundance.  Therefore, using line ratios that rely on \oiii$\lambda\lambda4959,5007$ as a diagnostic of ionization parameter will be unreliable at metallicities where these lines are weak.

In this work, we use the \hii\ regions of the Local Group galaxy M33 to explore the regime where \oiii$\lambda\lambda4959,5007$ is, and is not, effective as a diagnostic of ionization parameter for the purpose of evaluating radiation feedback and LyC escape.  In order to generate the emission-line images required for this analysis, it is necessary to carry out continuum subtraction, and we also further explore this process.

\section{Observations of M33} \label{sec:obs}

The north half of M33 was observed with the MOSAIC-1.1 imaging camera at the Mayall 4-m telescope, Kitt Peak National Observatory, on 2011 October 28--29.  We used the narrowband filters for \oii$\lambda3727$ (``O2", FWHM 50\AA), \oiii$\lambda5007$ (``O3", FWHM 50\AA), and \sii $\lambda6724$ (``ha16", FWHM 81\AA) for line imaging. For the continuum, we used broadband filters BATC454 ($\sim 4320 - 4700$ \AA) and BATC705 ($\sim 6950 - 7920$ \AA) for continuum subtraction of \oii\ and \oiii; and of \Halpha\ and \sii, respectively.  The continuum filters are used with the kind permission of R. Windhorst.  We also used archive \Halpha\ observations obtained in 2001 with the same setup by \citet{Massey2007AJ....134.2474M}.  There were $6\times 1400$ s, $4\times 600$ s, $5\times600$ s $+1\times900$ s, and $5\times300$ s exposures in \oii, \oiii, \sii, and \Halpha, respectively; and $5\times500$ s and $2\times1050$ s $+3\times 550$ s exposures in the blue and red continuum filters, respectively.  All observations were dithered to cover the gaps between the eight MOSAIC CCD chips. After being median-combined, the images have a variety of residual defects, including edge effects generated by the image combining process and a very low-level reflection of the telescope pupil.

In the \sii\space image, the giant \hii\space regions NGC 604, NGC 595, IC 131 and the star BD $+$30 243 generated trails likely caused by CTE effects,
causing a band $\sim$1000 pixels wide running through one section of the image. The northeast corner of the \sii\ continuum image also has an unusually bright patch, possibly due to reflections. Such bad pixels were trimmed out of the final array when performing the continuum subtraction. 
We use routines from the \texttt{astropy photutils} package to identify and mask the foreground Milky Way stars\deleted{, and interpolate over the masked pixels.} \added{We then interpolate over these masked pixels using a bilinear interpolation so that the faint wings of foreground stars do not affect the flux measurement of \hii\ regions.}\explain{Addressing minor comment 5 of the referee.}

We use a sample of 7 objects from \citet{TsC2016MNRAS.458.1866T}, listed in Table \ref{tab:TSC-dered-rered-fluxes}, to carry out flux calibration. 
For \sii\ and \oii, we sum the reported flux from the two lines forming these doublets, which are both included in the filter bandpasses.
We recover the un-dereddened fluxes using the reported $c(\rm H\beta)$ and extinction law: 
\begin{equation} \label{reddening law}
    \log\left(F_{\lambda_{1}} / F_{\lambda_{2}}\right)_{\text{corr}} = \log\left(F_{\lambda_{1}} / F_{\lambda_{2}}\right)_{\text{obs}} + C \left(f_{\lambda_{1}} - f_{\lambda_{2}}\right)
\end{equation}
The values in Table \ref{tab:TSC-dered-rered-fluxes}, columns (6)-(9) are the un-dereddened fluxes ($F_{obs}$) that we adopt for calibration. 

\begin{deluxetable*}{cccccccccc}
\tablenum{1}
\tablecaption{Line fluxes after removing reddening correction\tablenotemark{a}  \label{tab:TSC-dered-rered-fluxes}\explain{Added uncertainty for literature ratios}}
\tablewidth{0pt}
\tablehead{
\colhead{ID} & \colhead{c(\Hbeta)\tablenotemark{b}} & \multicolumn4c{Reddened} & \multicolumn2c{\oiii/\Halpha\ literature} & \multicolumn2c{\oiii/\Halpha\ Observed}\\
 & & \colhead{\Halpha} & \colhead{\oii $\lambda 3727$} & \colhead{\oiii $\lambda 5007$} & \colhead{\sii $\lambda\lambda 6718, 6732$} & mean & st.d. & mean & st.d.
}
\decimalcolnumbers
\startdata
B2011 b5 & 0.67 & 460 & 89 & 194 & 56 & 0.42 & 0.10 & 0.34 & 0.07 \\
IC 131 & 0.51 & 402 & 116 & 332 & 60 & 0.83 & 0.20 & 1.06 & 0.15 \\
BCLMP 290 & 0.12 & 318 & 188 & 177 & 32 & 0.56 & 0.07 & 0.49 & 0.05 \\
NGC 588 & 0.16 & 315 & 90 & 475 & 19 & 1.51 & 0.24 & 1.25 & 0.06 \\
BCLMP 626 & 0.02 & 292 & 239 & 165 & 43 & 0.57 & 0.06 & 0.66 & 0.06 \\
LGC HII3 & 0.09 & 302 & 171 & 285 & 31 & 0.94 & 0.12 & 1.03 & 0.29 \\
IC 132 & 0.37 & 370 & 46 & 566 & 15 & 1.53 & 0.20 & 1.89 & 0.18 \\
\enddata
\tablenotetext{a}{All fluxes are relative to \Hbeta = 100.}
\tablenotetext{b}{c(\Hbeta) values from \citet{TsC2016MNRAS.458.1866T} used to remove their reported reddening correction.}
\end{deluxetable*}

We apply rectangular apertures corresponding to the reported slit width at the position for each object given in 
Table 1 of \citet{TsC2016MNRAS.458.1866T}. We use the \texttt{photutils} aperture photometry routine in \texttt{astropy} to obtain the photometry.
To account for positional inaccuracies arising from atmospheric seeing, we offset the slit positions by a normally distributed random variable with standard deviation equal to the seeing in pixels. We average over 20 such randomly offset apertures for the integrated photometry. 
The computed ratios are then used together with the calibrated flux ratios derived from the data of \citet{TsC2016MNRAS.458.1866T}
to determine the flux calibration, using the \Halpha/\Hbeta\ ratio of 2.86.

\section{Continuum subtraction} \label{sec:subtract}
Narrow-band imaging data include flux from both line emission and diffuse stellar continuum. To isolate the line flux, an off-line image containing only the continuum is usually subtracted. Due to differing filter transmissions and variation in the continuum spectral energy distribution (SED), the continuum image must be scaled before subtraction, and determining the scale factor is nontrivial. \citet{Hayes2009AJ....138..911H} and \citet{James2016ApJ...816...40J} have utilised stellar population synthesis modeling that compute spatially varying scale factors on a pixel-by-pixel basis. These methods are model-dependent and \citet{Hayes2009AJ....138..911H} discuss their advantages and pitfalls in detail. 

In this work, we focus on empirical methods that compute a single characteristic scale factor for a large region or an entire image. 
\citet{Keenan2017ApJ...848...12K} describe a method where a single optimal scale factor is found for such a region. They note a slope change in the mode of the pixel values vs scale factor for the continuum-subtracted images. \citet{Keenan2017ApJ...848...12K} show that this transition results when the scale factor induces any oversubtraction. At small values of the scale factor, the mode of the pixel-value histogram is determined by the lowest-value pixels. The change in flux for these pixels is small as the scale factor varies, hence the slope of mode vs scale factor is shallow. At higher values of the scale factor, the mode is dominated by oversubtracted pixels.  As the scale factor increases, the first pixels to become oversubtracted are the brightest ones, and their flux has a strong dependence on scale factor. Thus, the slope of the mode vs scale factor is steeper in the over-subtracted regime (see \citet{Keenan2017ApJ...848...12K} for more details).

\citet{Hong2014PASP..126...79H} present a method that similarly identifies the transition to oversubtraction, but based on the skewness of the pixel-value histogram as a function of the scale factor. The above two methods have been shown to work well for images where there are a significant number of continuum-dominated pixels. \deleted{However, we still encountered some difficulties in applying both of these.} 
\added{However, we still encountered some difficulties in adequately constraining the best scale factors.
In particular, the slope transition reported by \citet{Keenan2017ApJ...848...12K} can be hard to discern. There can be multiple slope changes and oscillatory behavior in the mode, as \citet{Keenan2017ApJ...848...12K} indicate. The exact point of transition is therefore uncertain.
Similarly, the method used by \citet{Hong2014PASP..126...79H} relies on computing the second derivative of the skewness with respect to the scale factor. Estimating this derivative accurately requires a very fine search through the scale factor space.  
Appendix~A demonstrates the functionality of these two methods (Figures \ref{fig:haro-11-fraction}, \ref{fig:eso338-fraction} and \ref{fig:mrk71-fraction}). 
}
In this paper, we therefore propose a revised method that uses the mode to obtain the optimal scale factor, 
\added{which can provide a narrower confidence interval while also reducing the computing power needed}.
\explain{Addressing comment 1.2 of referee}

\subsection{Revised Mode Method to Identify Scale Factor}  \label{subsec:agg-factor}

The line emission represents an excess signal over the continuum, hence, the pixel-value distributions will have a tail to positive values dominated by the real signal.  Since we are interested in identifying the diffuse background to carry out the continuum subtraction, we therefore employ a $3\sigma$ filter from the \texttt{scipy} library on the line image. The routine computes the mean and standard deviation $\sigma$ of the data, then rejects any data points that are $>3\sigma$ from the mean. The mean and standard deviation are recomputed and the filtering is done again. This iterative process is continued until there are no more rejections. At each scale factor, we first subtract the scaled continuum from the line image, then invoke the $3\sigma$ filter on the subtracted image. This removes much of the real signal, leaving a residual histogram that is more dominated by the diffuse continuum signal. The resulting pixel-value distribution is better suited for the purpose of identifying the optimal continuum image scale factor.

\begin{figure*}
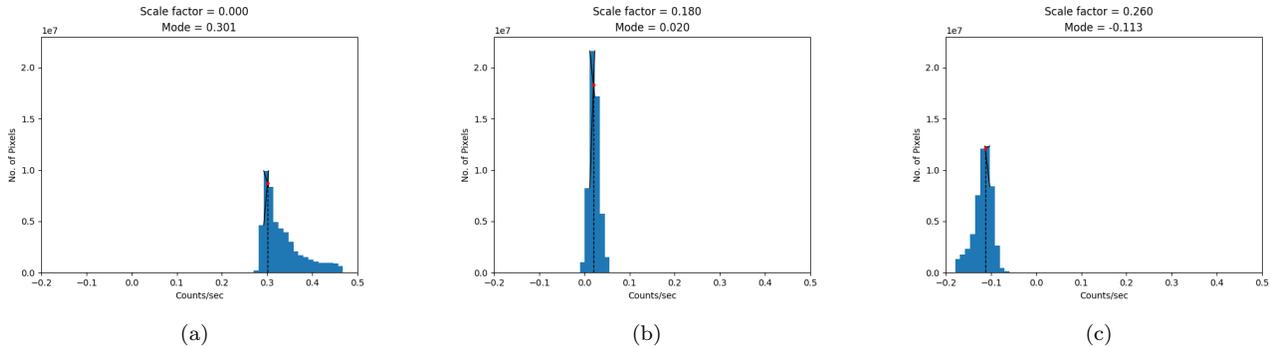

\gridline{\fig{M33_factor_0.0000.png}{0.3\textwidth}{(a)}
          \fig{M33_factor_0.1800.png}{0.3\textwidth}{(b)}
          \fig{M33_factor_0.2600.png}{0.3\textwidth}{(c)}
          }
\caption{Histograms for M33 \oiii\ continuum subtraction. Panel (a) shows undersubtracted pixel histogram with high-value tail. Panel (b) shows the optimally subtracted pixel histogram, where the number of pixels in the 3 modal bins is maximized.  Panel (c) shows the emergence of the low-value tail in an oversubtracted pixel histogram. }
\label{fig:histogram}
\end{figure*}

Similar to \citet{Hong2014PASP..126...79H}, we observe a transition in the skewness of the pixel histogram as the scale factor is increased. 
Figure~\ref{fig:histogram} shows how the shape of the pixel-value distribution changes as the scale factor is varied. At low scale factors (Figure \ref{fig:histogram}(a)), the image is undersubtracted. The distribution for the subtracted image resembles the initial flux distribution, where pixels with high continuum values contribute to high-value bins. As a result, the histogram is skewed to the right.

As the scale factor is increased, the flux distribution in the line image approaches the flux distribution in the scaled off-line image. 

Ideally, at this point of optimal continuum subtraction, the background flux should be zero. This is characterised by the mode of the pixel histogram being zero. However, this is not always the case, since the sky may have a different SED compared to the diffuse stellar continuum. 
This issue is  particularly relevant for ground-based observations where the sky background is significant. \added{Once the scale factor has been set by subtracting the stellar emission, the residual sky background can be eliminated while performing aperture photometry, as we have done for the flux measurements in Tables \ref{tab:TSC-dered-rered-fluxes} and \ref{tab:object-list}}
\explain{Addressing minor comment 1 of referee.}

On further increasing the scale factor, the pixels with strong continuum flux now contribute to the negative bins as they are the first ones to become oversubtracted. The spread increases in the negative direction, skewing the histogram to the left (Figure \ref{fig:histogram}(c)). This is a reversal of the behavior seen earlier, and it corresponds to the transition in skewness reported by \citet{Hong2014PASP..126...79H}. It should be noted that the negative tail is less statistically `heavy' compared to the positive tail, due to the continued presence of pixels with emission-line flux. 
\citet{Hong2014PASP..126...79H} demonstrate the same effect as a slight positive bias to the skew at the transition point.

\begin{figure*}
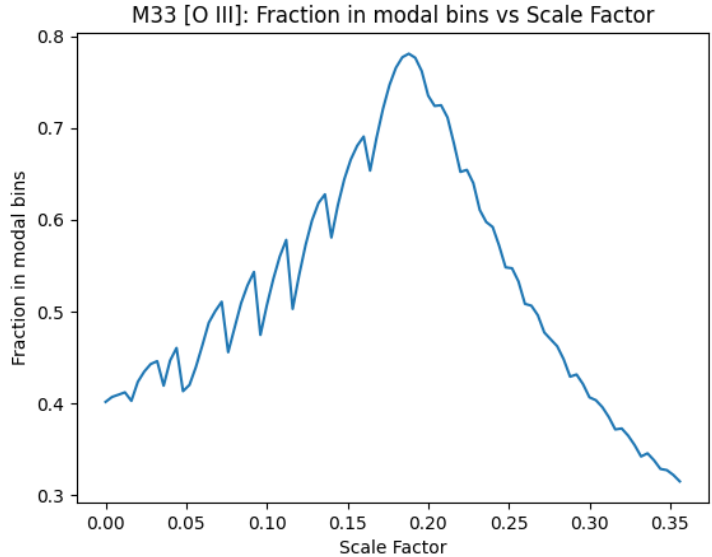

    \fig{M33_OIII_Fraction_in_modal_bins.png}{0.6\textwidth}{}
    \caption{Modal bin fraction, calculated from equation \ref{modelines}, vs the scale factor. This metric is maximised in this example when the scale factor is 0.188, indicating the point of optimal subtraction.  The scalloped pattern is caused by the binning of the pixel value distribution (see text).
    \label{fig:fraction-chart}}
\end{figure*}

At the optimal scale factor, the number of background pixels is therefore maximized in the modal bin, as shown in Figure \ref{fig:fraction-chart}.
Due to the background noise, the optimally subtracted histogram will still have some spread around the mode. We therefore also consider the total number of pixels in the bins adjacent to the modal bin. This is also helpful if the mode falls on the boundary between two bins. Using all three bins provides some robustness against such cases.  Thus, the optimal scale factor is that which maximises $({f_0 + f_1 + f_2})/{N}$, where $f_0$ is the number of pixels in the modal bin; $f_1$ and $f_2$ are the number of pixels in the pre-modal and post-modal bins, respectively; and $N$ is the total number of pixels in the image, after applying the $3\sigma$ filter. If the modal bin is the first or last bin, then $f_1$ or $f_2$ are accordingly assumed to be zero. The value of $N$ changes with the scale factor due to the $3\sigma$ filter. Initially, the brightest pixels get rejected, but at the correct scale factor, the flux from these correctly subtracted pixels falls within the $3\sigma$ limit. 
\added{The chosen metric, $({f_0 + f_1 + f_2})/{N}$, is computed using only the pixel histogram at the current scale factor. In comparison, the methods of \citet{Keenan2017ApJ...848...12K} and \citet{Hong2014PASP..126...79H} require the histograms at adjacent values of the scale factor to compute the slope of the mode, or the second derivative of skewness which requires finer sampling to reduce the error bound.}
\explain{Addressing major comment 1 of the referee.}

A jagged or scalloped pattern is apparent in Figure \ref{fig:fraction-chart} as the scale factor is increased. The pattern arises due to the binning criteria chosen for the of the modal bin fraction. In particular, it is due to the choice of using only the 3 modal bins to measure of the peak of the pixel distribution. In the case of highly skewed histograms, the peak is inadequately sampled by only 3 bins, and small changes in the pixel distribution are amplified, giving rise to the jagged pattern seen. This behaviour can be smoothed by modifying the criterion to include more bins, or by finer sampling of the pixel distribution by increasing the number of bins. Finer sampling has an added computational cost. For the purposes of identifying the globally optimal scale factor from a symmetric histogram, the chosen metric works well and is unaffected by the local variations due to skewed histograms.
\added{Appendix~A demonstrates this method, with comparisons to the \citet{Keenan2017ApJ...848...12K} and \citet{Hong2014PASP..126...79H} methods (Figures~\ref{fig:haro-11-fraction}, \ref{fig:eso338-fraction} and \ref{fig:mrk71-fraction}).}
\explain{Addressing major comment 1}

We caution that the modal bin maximisation works to subtract the dominant background component in the image, regardless of whether it is diffuse starlight, sky emission, or diffuse nebular emission.  Thus, if the goal is to subtract diffuse starlight, then that component should constitute a significant fraction of the total pixel population. 
For example, if sky pixels dominate the variance, then the algorithm produces an undersubtracted image relative to diffuse starlight. This can be caused by two effects. First, if sky emission dominates the flux of the background pixels, then the algorithm identifies the optimal scale factor for the sky background. This is illustrated in Appendix \ref{sec:other-applications}. Similarly, if the random variance of the sky pixels is high in images with low signal-to-noise, then the large variance makes the reduced spread of the stellar continuum pixels harder to detect. This can be explained as follows: the unsubtracted line image has a certain variance arising from the true signal, which is gradually subtracted out by increasing the scale factor. At the same time, we introduce additiional variance through the random noise of the sky pixels, which increases on increasing the scale factor. If the signal-to-noise ratio is low, the reduction in variance of diffuse starlight is drowned out by the noise that we introduce. The spread is minimised at a lower scale factor, resulting in an undersubtracted image. We address this issue in Section \ref{subsec:application} and Appendix \ref{sec:other-applications}.

On the other hand, the presence of widespread diffuse line emission will also bias the pixel-value histogram and the spread towards higher values. Contamination of pure continuum pixels with diffuse line emission therefore also inflates the variance of continuum pixels. The algorithm proposed above 
overcompensates for the larger spread, resulting in a scale factor greater than optimum. This is mitigated by ensuring the presence of pure continuum pixels in the image. Therefore, prior knowledge about the emission region characteristics is required. In short, the background must be dominated by diffuse starlight pixels in order for these statistical methods to identify the optimal scale factor to subtract this background component. 
\added{See \citet{Hong2014PASP..126...79H} for a discussion on how the various background compositions affect these continuum-subtraction methods.}
\explain{Addressing minor comment 1.}


\subsection{Application of the Method}\label{subsec:application}

Like most spiral galaxies, M33 has a strong color gradient in its diffuse starlight, implying that the scale factor for subtracting this component will vary spatially. We therefore define 5 regions using elliptical isophotes (Figure \ref{fig:el-offb_img}), which are are chosen by visual inspection of the approximate stellar surface brightness. Region 0, the galactic center, is dominated by resolved sources and does not have many background pixels. 
When presented with an image like this, our algorithm tends to produce oversubtracted images since there is not enough background to determine the correct scale factor. The opposite holds for Region 4, which is dominated by sky pixels and relatively lacking in diffuse starlight. Sky-dominated images tend to undersubtract diffuse starlight when we apply our method, as described in Section~\ref{subsec:agg-factor}. Regions 1, 2 and 3 have a good mix of emission, continuum, and background pixels, so our method works well for such regions.  Due to the different characteristics of the pixel populations of Regions 0 and 4, we assign 
the scale factors for Region 0 and 4 by setting the modes of their continuum-subtracted pixel values equal to the modes for Regions 1 and 3, respectively.

\begin{figure*}
    \begin{interactive}{animation}{M33_OIII_movie.mp4}
        \plotone{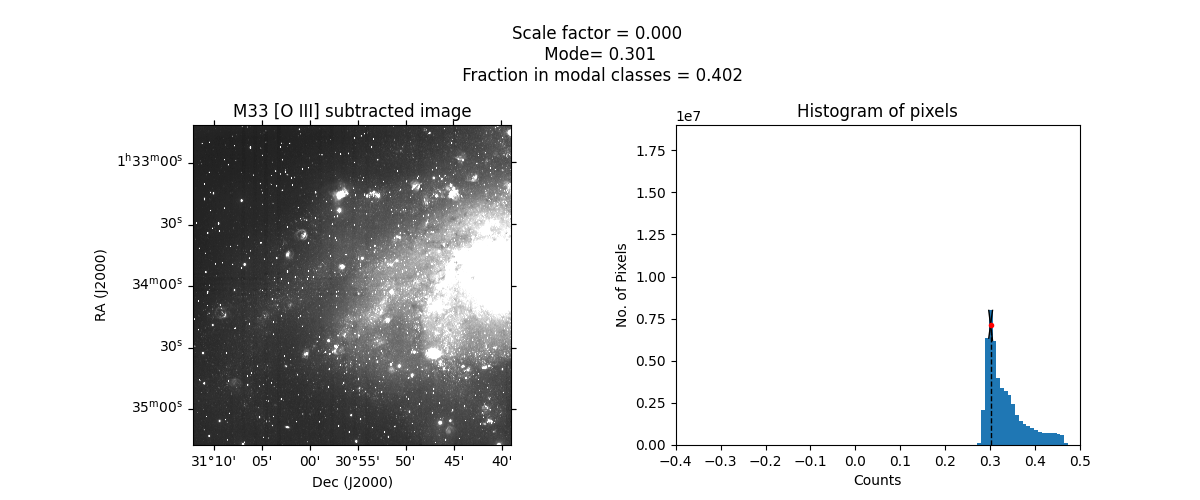}
    \end{interactive}
    \caption{The online version of this figure is an animation of the step-by-step subtraction process, showing how the shape of the pixel histogram changes for the corresponding continuum-subtracted \oiii\ images as the continuum-image scale factor is increased from 0 to 0.36. The point of optimal subtraction can be identified around scale factor $\approx 0.18$. The video duration is 9 seconds. 
    \label{fig:video}}
\end{figure*}

In our pipeline, we first employ the $3\sigma$ filtering described above to the emission-line image. The pixels with the strongest fluxes are rejected by the filter.
Next, we generate the pixel-value histogram of the continuum-subtracted image. The bin width is chosen following \citet{Sturges1926}: $\text{The number of bins}\ n \approx 1 + \log_{2}N$, where $N$ is the total number of pixels. For our images, this value is 22-25 bins, depending on the Region. The modal bin is identified and the modal bin fraction calculated as described earlier in Section \ref{subsec:agg-factor}. The exact value of the mode $M$  can be calculated by taking the intersection of two lines that linearly interpolate the data in the modal bin:
\begin{equation}\label{modelines}
M = L + \frac{f_{0} - f_{1}}{2f_{0} - f_{1} - f_{2}}\times h 
\end{equation}
where $L$ is the lower limit of the modal bin, and $h$ is the bin width. This exact value is needed to set the scale factors for Regions 0 and 4.
An animation of the continuum subtraction process (Figure~\ref{fig:video}) is available in the online version of this paper. 
\added{Table~\ref{tab:object-list} gives our narrowband fluxes for the objects, with
errors estimated by computing the median background around each object during aperture photometry. We confirmed the \Halpha\ luminosities of three giant \hii\ regions: NGC 604, NGC 595 and IC 131 with the values given by \citet{Relano2009ApJ...699.1125R}.}
\explain{Addressing minor comment 4.2 by referee.}

\begin{figure*}
\gridline{\fig{M33_oiii_el.png}{0.5\textwidth}{(a)}
          \fig{M33_oiii_offb.png}{0.5\textwidth}{(b)}
          }
\caption{Narrow-band \oiii\ on-line (a) and off-line (b) images of M33.  The five regions with varying background levels are shown.
\label{fig:el-offb_img}}
\end{figure*}

To further test our new method of continuum subtraction, we also apply it to \ciii\ $\lambda1909$ narrow-band imaging of three starburst galaxies, Haro 11 \citep{Micheva2020ApJ...903..123M}, ESO338-IG04, and Mrk 71. Whereas M33 has a strong color gradient and many resolved individual stars, these more distant galaxies are dominated by more uniform, diffuse starlight.  These examples also have very faint line emission compared to the M33 data.  We find that our new continuum subtraction technique works well on these galaxies.  The results are shown in Appendix~\ref{sec:other-applications}.

\begin{deluxetable*}{cccccDDDDDDDD}
\tablenum{2}
\tablecaption{\hii\space Regions in Northern M33 \label{tab:object-list}}
\tablewidth{\linewidth}
\tabletypesize{\scriptsize}
\tablehead{
\colhead{Object} & \colhead{RA (J2000)} & \colhead{Dec (J2000)} &
\colhead{12+log(O/H)\tablenotemark{a}} & \colhead{Opacity\tablenotemark{b}} & 
\multicolumn2c{\Halpha\tablenotemark{c}}   &  \multicolumn2c{\Halpha\ err\tablenotemark{c}}   &
\multicolumn2c{\oii\tablenotemark{c}}   &  \multicolumn2c{\oii\ err\tablenotemark{c}}   &
\multicolumn2c{\oiii\tablenotemark{c}}   &  \multicolumn2c{\oiii\ err\tablenotemark{c}}   &
\multicolumn2c{\sii\tablenotemark{c}}   &  \multicolumn2c{\sii\ err\tablenotemark{c}}  
}
\decimalcolnumbers
\startdata
BCLMP 616  & 1h32m54.38s & 30d50m28s   & $8.226^{+0.006}_{-0.005}$             & 1       & 14.6 & 0.22     & 7.61  & 0.20      & 2.03   & 0.46            & 2.97  & 0.40           \\
LHK2017 54 & 1h32m56.28s & 30d40m36.7s & $8.383^{+0.004}_{-0.003}$             & 4       & 23.6 & 3.9       & 13.1  & 1.2        & 5.2   & 2.6             & 13.1  & 4.5            \\
BCLMP 289  & 1h32m57.5s  & 30d44m27s   & $8.208^{+0.003}_{-0.004}$             & 3       & 57.4 & 19     & 31.4  & 0.83           & 66.4   & 4.6             & 18.1  & 21            \\
CPSDP 67   & 1h32m59.21s & 30d41m20.8s & $8.518^{+0.007}_{-0.007}$             & 1       & 10.1 & 1.4     & 4.91  & 0.16            & 2.65   & 1.4             & 3.64  & 2.0            \\
BCLMP 285  & 1h33m02.88s & 30d41m08.1s & $8.238^{+0.003}_{-0.003}$             & 1       & 13.1 & 0.25     & 5.95  & 0.04            & 4.63   & 0.27             & 2.13  & 0.42
\enddata
\tablenotetext{a}{From \citet{Lin2017ApJ...842...97L}.}
\tablenotetext{b}{$1=$ optically thick, $2=$ blister, $3=$ optically thin, $4=$ shock, $5=$ indeterminate.}
\tablenotetext{c}{Luminosities given in $10^{36}$ erg/s.}
\tablecomments{Data for first five objects are shown here. The online version of this paper has data for all 108 objects in a machine readable format.}
\end{deluxetable*}

\section{Ionization-parameter mapping}\label{sec:IPM}

We generate line ratio maps for \oiii/\oii\ and \oiii/\sii\ as described by \citet{Keenan2017ApJ...848...12K} and \citet{Pellegrini2012ApJ...755...40P} to evaluate the optical depth of photoionized \hii\ regions by ionization-parameter mapping (IPM).
IPM is most directly applied to distinct objects, and significant diffuse continuum can mask the signal from individual HII regions. Unsharp masking allows us to locally remove the average diffuse ionized gas emission over a given length scale. This allows for an amplification in the signal for optically thin nebulae, where the low-ionization emission may be dominated by diffuse ambient emission. The residual presence of field stars affects the median smoothing, so we first 
\added{carry out a bilinear interpolation}
\explain{addressing minor comment 5 of the referee}
over these across regions three times the stellar PSF. We then mask any emission with intensity greater than 30 counts ($5.8\times 10^{-15}\ \rm erg\ s^{-1}\ cm^{-2}\ px^{-1}$). \added{This is approximately the saturation level of the detector.}\explain{Addressing minor comment 7 of the referee.} We median filter the remaining emission over a length scale 30$\times$ the stellar PSF, corresponding to 83.1 pixels.
Finally, we subtract that medianed image from the original data, smoothed to the seeing to improve S/N.

Following \citet{Pellegrini2012ApJ...755...40P}, we classify the M33 \hii\ regions into 5 classes: (0) indeterminate, (1) optically thick, (2) blister, (3) optically thin, and (4) shocked, based on the ionization structure in the halos of individual regions. Objects are considered optically thick if \oii\ and/or \sii\ dominates over at least two-thirds of the circumference in projection.  Blister and optically thin \hii\ regions are those with the low-ionization species dominating over one-third to two-thirds; and less than one-third of the circumference, respectively.
Shocked nebulae show an ionization structure that is unlikely to have resulted from photoionization alone, with a pocket of low ionization species on the interior surrounded by \oiii\ on the outside. Objects are categorised as ``indeterminate" due to poor S/N, incomplete data, or abnormal ionization morphology. 

\begin{figure*}
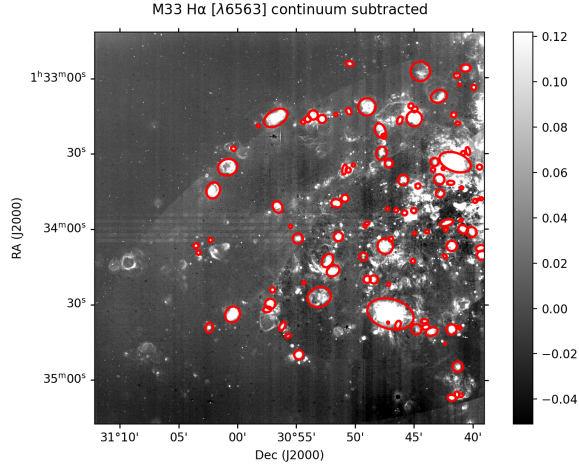

  \fig{M33_Ha_finder.png}{0.5\textwidth}{}
  \caption{Continuum-subtracted \Halpha\ image with the sample \hii\space regions from Table \ref{tab:object-list} marked.
  \label{fig:finder-chart}}
\end{figure*}

\begin{figure*}
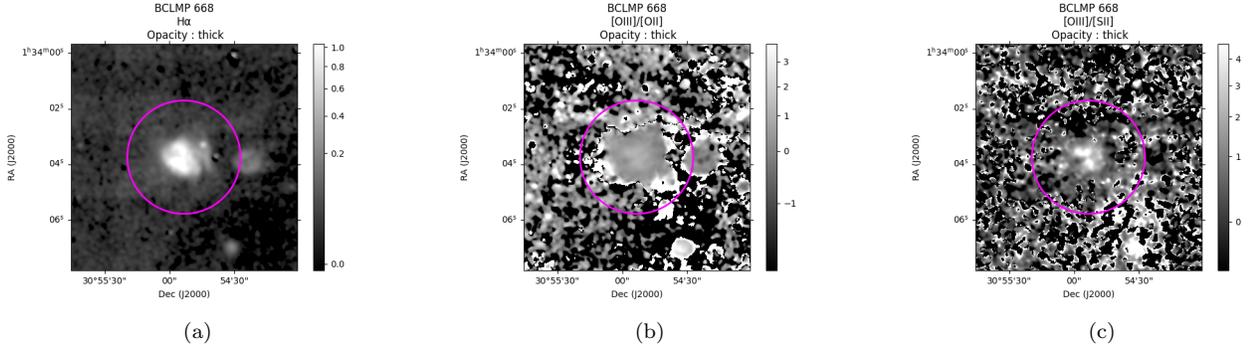

\gridline{\fig{BCLMP_668_halpb.png}{0.3\textwidth}{(a)}
          \fig{BCLMP_668_o3o2b.png}{0.3\textwidth}{(b)}
          \fig{BCLMP_668_o3s2b.png}{0.3\textwidth}{(c)}
          }
\caption{BCLMP 668 in smoothed, unsharp-masked (a) continuum subtracted \Halpha, (b) \oiii/\oii, and (c) \oiii/\sii\space ratio maps.  
\added{Color bar units are in image counts.}
This object is classified as optically thick, and displays a shell-like structure of high vs low ionization zones.}
\label{fig:bclmp-668}
\end{figure*}

\begin{figure*}
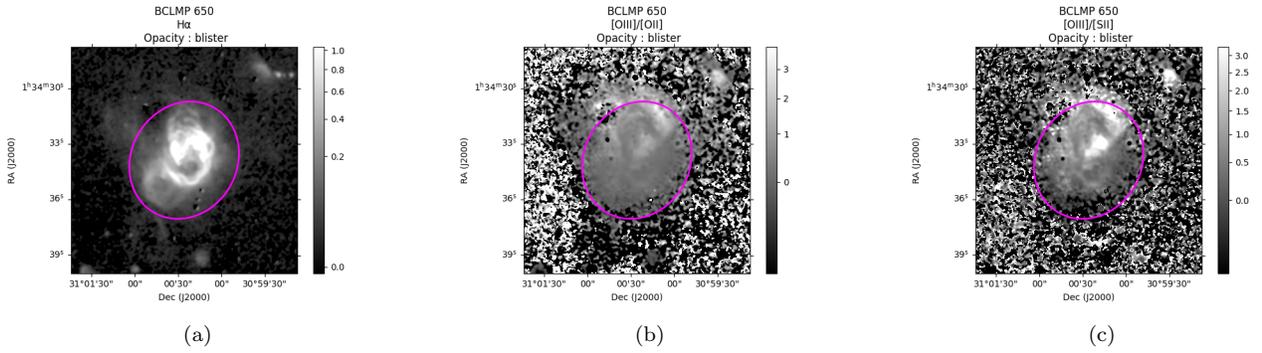

\gridline{\fig{BCLMP_650_halpb.png}{0.3\textwidth}{(a)}
          \fig{BCLMP_650_o3o2b.png}{0.3\textwidth}{(b)}
          \fig{BCLMP_650_o3s2b.png}{0.3\textwidth}{(c)}
          }
\caption{Same as Figure~\ref{fig:bclmp-668} for BCLMP 650.  This object's optical depth is classified as blister, with highly ionized gas subtending between one-third and two-thirds of the object's circumference.}
\label{fig:bclmp-650}
\end{figure*}

\begin{figure*}
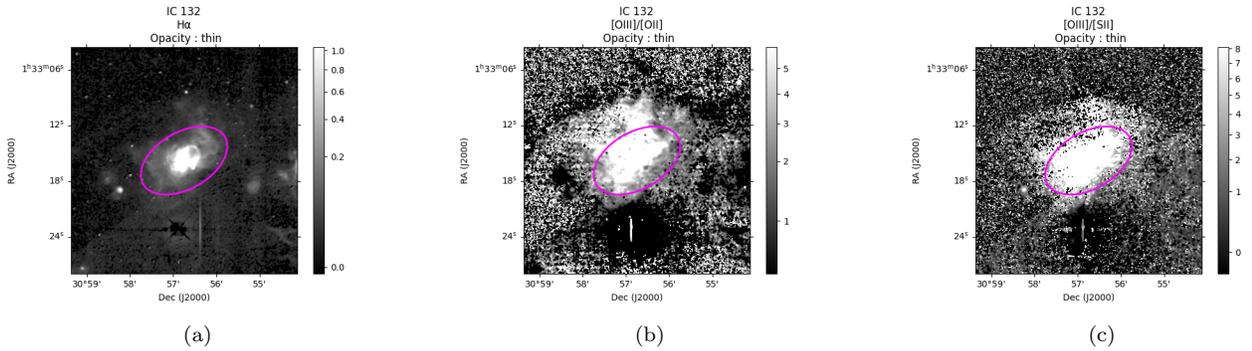

\gridline{\fig{IC_132_halpb.png}{0.3\textwidth}{(a)}
          \fig{IC_132_o3o2b.png}{0.3\textwidth}{(b)}
          \fig{IC_132_o3s2b.png}{0.3\textwidth}{(c)}
          }
\caption{Same as Figure~\ref{fig:bclmp-668} for IC 132. Highly ionized \oiii\space is seen all around the halo, indicating the escape of ionizing radiation.}
\label{fig:ic-132}
\end{figure*}

\begin{figure*}
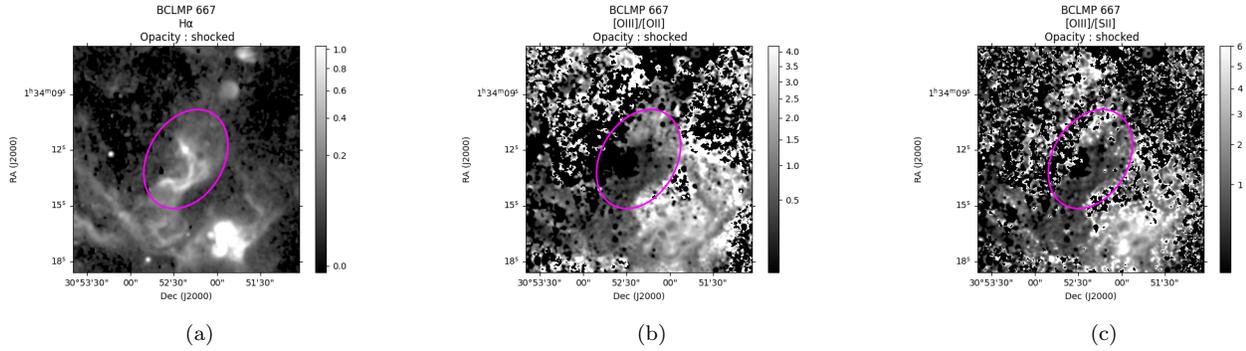

\gridline{\fig{BCLMP_667_halpb.png}{0.3\textwidth}{(a)}
          \fig{BCLMP_667_o3o2b.png}{0.3\textwidth}{(b)}
          \fig{BCLMP_667_o3s2b.png}{0.3\textwidth}{(c)}
          }
\caption{Same as Figure~\ref{fig:bclmp-668} for BCLMP 667.  This object is classified as shocked, and the ratio maps reveal the inverted ionization profile.}
\label{fig:bclmp-667}
\end{figure*}

Our classifications are given in Table~\ref{tab:object-list} for our sample of 108 objects, and are based on consideration of both \oiii/\sii\ and \oiii/\oii\ ratio maps. The object locations in the disc of M33 are shown in Figure \ref{fig:finder-chart}. We present examples of categories  1--4 in Figures~\ref{fig:bclmp-668} to \ref{fig:bclmp-667}. These figures show the \oiii/\sii\ and \oiii/\oii\ ratio map for these representative objects.
Similar images for all the objects in our sample are provided in the interactive version of Figure~\ref{fig:metal-chart} below.

Figure \ref{fig:bclmp-668} shows ratio maps for BCLMP 668, an optically thick nebula. As evidenced by Figure \ref{fig:bclmp-668}(b) and (c), 
this object shows a classic, Str\"omgren sphere structure, with the low-ionization envelope completely surrounding the highly ionized core.
The emission from \oiii \space is relatively low throughout.

Figure \ref{fig:bclmp-650} shows ratio maps for BCLMP 650, a blister object. In Figure \ref{fig:bclmp-650}(b) and (c), we see that the envelope of the low-ionization species \oii\space and \sii\space extend about halfway around the \hii\space region in projection. There is a break to the west (upper portion in image), where \oiii\space dominates, allowing the escape of ionizing radiation.

IC 132 (Figure \ref{fig:ic-132}) is an optically thin \hii\space region that is extremely bright in \oiii. Figure \ref{fig:ic-132}(b) shows that the ratio of \oiii\wavelength{5007}/\oii\wavelength{3727}  is $>1$ over the entire visible extent of the object.
The ionization structure in Figure \ref{fig:ic-132}(b) is similar to that found by \citet{LopezHernandez2013MNRAS.430..472L} for this object. Comparing Figure \ref{fig:ic-132}(c) with other objects demonstrates how the \oiii\wavelength{5007}/\sii\wavelength{6724} \space morphology changes dramatically for optically thick vs thin regions.

The morphology of BCLMP 667 (Figure \ref{fig:bclmp-667})
shows the lower-ionization species dominating along the inside curve of this object, which is the opposite of what is expected for photoionization from a source driving the shell-like nebular morphology. We therefore classify this as a shock-ionized object.

\subsection{Metallicity dependence of IPM}\label{subsec:metal-dependence}

Figure \ref{fig:metal-chart}(a) shows \oiii/\oii\ excitation vs $12+\log(\rm O/H)$, with optical depth class shown by the symbol colors.
We have used data from \citet{Lin2017ApJ...842...97L} for the values of $12 + \log\rm (O/H)$ for the  \hii\space regions in our sample (Table \ref{tab:object-list}). As expected, Figure \ref{fig:metal-chart}(a) shows a strong anticorrelation between optical depth and \oiii/\oii\ ratio.  However, there is also a clear trend with metallicity, and there are no objects classified as optically thin (as opposed to blister) having $12 + \log\rm (O/H)> 8.25$. 

The strength of the \oiii\wavelength{5007} line exhibits a strong, non-linear relationship with the O abundance: it gets stronger with increasing metallicity in the metal-poor regime, but at higher metallicities, the greater abundance of metals cools the nebula, rendering it unable to collisionally excite \oiii\space in the visible-wavelength transitions. Thus, fine-structure lines in the infrared dominate the \oiii\ emission at higher $12+\log(\rm O/H)$. 
Bright-line abundance indices such as $R_{23}\equiv \rm ($\oii$\lambda3727 + \rm $\oiii$\lambda\lambda4959,5007)/\rm H\beta$ and O3N2 $\equiv \log((\rm$ \oiii$\lambda5007/\rm H\beta)/($\nii$\lambda6583/\rm H\alpha$)) are based on this principle, and are thus maximised in the interval $12 + \log\rm (O/H) = 8.0 \text{ to } 8.5$, dropping off steeply at higher values \citep[e.g.,][]{Yin2007A&A...462..535Y, Kewley2002ApJS..142...35K}.  For our sample,
\citet{Lin2017ApJ...842...97L} derive the values of $12+\log(\rm O/H)$ using the bright-line indices O3N2, N2 $\equiv \log(\rm [NII]6583/H\alpha)$
\citep[e.g.,][]{Marino2013A&A...559A.114M} as well as direct modelling of the electron temperature $T_{e}$, with a majority derived using methods reliant on the \oiii\wavelength{5007} emission line. We caution that \hii\ structural evolution effects can shift strong line ratios like N2 by more than an order of magnitude \citep{Pellegrini2020aMNRAS.496..339P}; however, the dominant trends with metallicity are well established.

Since we utilise the \oiii/\oii\ and \oiii/\sii\ ratios as a tracer for degree of ionization, the method of IPM used here is therefore also sensitive to the metallicity.  The optical \oiii\ lines are weak at higher metallicity even though the O$^{++}$ ion may still be prevalent, and therefore the efficacy of IPM is reduced in this regime. Thus, some of the optically thick objects may be mis-classifications on account of \oiii/\sii\ and \oiii/\oii\ being reduced at higher abundances. 

\begin{figure*}
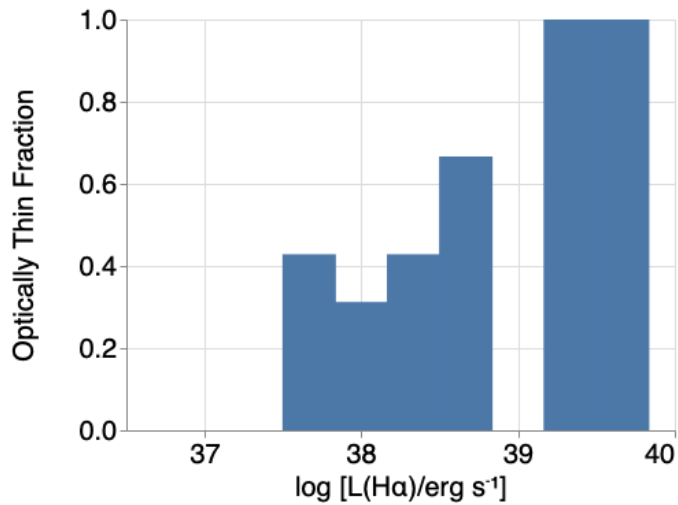

    \fig{thin_fraction.png}{0.5\textwidth}{}
    \caption{Fraction of optically thin \hii\space regions as a function of luminosity for $12+\log(\rm O/H) < 8.4$.  Luminosities of the sample objects range from $\log L(\rm H\alpha) = 36.7$ to  39.7, thus the frequencies in bins with $\log L(\rm H\alpha) < 37.5 = 0.$
    There are no objects in our sample that have $\log L(\rm H\alpha)$ in the range 38.8 to 39.2. }
    \label{fig:thinfraction}
\end{figure*}

As seen in the LMC and SMC, the most luminous \hii\ regions are also the most likely to be optically thin, including blister objects \citep{Pellegrini2012ApJ...755...40P}.  This trend is also seen in our sample in Figure~\ref{fig:thinfraction}, which shows the frequency of optically thin and blister \hii\ regions as a function of \Halpha\ luminosity for objects having $12+\log(\rm O/H) < 8.4$. 
In Figure~\ref{fig:metal-chart}(b), we see that the most optically thin objects are tightly clustered at the lowest metallicities.  Luminous \hii\ regions with $L(\rm H\alpha) \geq 10^{38}\ \rm erg\ s^{-1}$ at moderate metallicities of $12+\log(\rm O/H)=8.3$ or 8.4 are mostly classified as optically thick, while the opposite is true at lower metallicity.  This further supports the likelihood that some of the higher-metallicity, high-luminosity objects classified as optically thick are actually optically thin.

On the other hand, a real trend in increased frequencies of optically thin nebulae must also exist for metal-poor environments.  Dust content decreases with metallicity, thereby decreasing the opacity to the Lyman continuum.  Also, larger star clusters generating luminous \hii\ regions tend to be more prevalent at lower metallicity, increasing the likelihood of powering objects with the earliest O stars, and enhancing the likelihood of optically thin \hii\ regions. Metal-poor OB atmospheres also tend to be hotter than at solar metallicity \citep[e.g.,][]{MaederMeynet2001A&A...373..555M, MartinsPalacios2021A&A...645A..67M}, driving higher nebular ionization parameters. Thus, without detailed modeling of the individual objects, it is impossible to clarify the locus of optically thin objects in Figure~\ref{fig:metal-chart}, but the IPM classifications in Table~\ref{tab:object-list} likely significantly underestimate the frequency of optically thin nebulae.

\begin{figure*}
    \begin{interactive}{js}{metal_chart.zip}
        \fig{metal_chart.png}{0.4\textwidth}{(a)}
        \fig{metal_luminosity.png}{0.4\textwidth}{(b)}
    \end{interactive}
    \caption{\oiii$\lambda5007/$\oii$\lambda3727$ (panel (a), left) and \Halpha\ luminosity (panel (b), right) vs oxygen abundance.  Symbol type and color show optical depth classifications.
    Objects classified as optically thin and blister have higher \oiii/\oii\ values, which is a function of metallicity. 
    \added{Median measurement errors for \oiii/\oii, $\log L$(\Halpha), and $12+\log(\rm O/H)$ are 0.06 dex, 0.018 dex, and 0.004 dex. Systematic errors are on the order of 20\% for \oiii/\oii\ and $L$(\Halpha), and 0.18 dex for $12+\log(\rm O/H)$.}
    An interactive version of panel (b) will be made available in the online version. Users can: \\ 
    i) Filter the objects by $\log$(\oiii/\oii) by means of a slider. \\
    ii) Select an object by clicking on the symbols or in a drop-down list to view the corresponding ratio maps used for classification as shown in Figures \ref{fig:bclmp-668} -- \ref{fig:bclmp-667}. Both monochrome and colour images are available. \\
    iii) Select the different optical depth categories in the legend to plot only objects in the selected categories. \\
    iv) Hover with a mouse pointer on the symbols to view the ID, metallicity and \oiii/\oii\ value for each object.}
    \label{fig:metal-chart}
\end{figure*}

The classifications are also dependent on the spatial resolution and depth of the ratio maps.  Comparing Figure~\ref{fig:thinfraction} with the results of \citet{Pellegrini2012ApJ...755...40P} for the LMC and SMC, the frequency of optically thin objects is much lower, about a factor of 2 at $L(\rm H\alpha)\ \sim10^{38}\ \rm erg\ s^{-1}$.  There is no reason to believe that the ISM properties of M33 are substantially different than in these galaxies, and so most likely our classifications are affected by the lower spatial resolution of the M33 imaging data (22 pc in the smoothed images) relative to the imaging of the Magellanic Clouds (1.4 pc).
\added{Given the qualitative similarity of the results from these two studies, the systematic errors generated by the resolution effects do not substantively change the observed trend in Figure~\ref{fig:thinfraction}.  However, we caution that absolute interpretation of optical depths is much less reliable.} 
\explain{Addressing referee comment 2.2}

We therefore conclude that IPM based on \oiii/\oii\ or \oiii/\sii\ is useful only at lower metallicities, $12 + \log (O/H) \lesssim 8.4$, and it is also dependent on spatial resolution.  Other diagnostic lines can extend the use of IPM.  For example, \citet{Zastrow2013ApJ...779...76Z} have similarly used \siii$\lambda 9069/\rm$\sii$\lambda6724$ ratio maps to identify Lyman continuum escape. In general, the underlying principle remains the same: to reveal the nebular ionization structure by differentiating the high vs low ionization zones.

\subsection{A kpc-sized patch of elevated \oiii}

\begin{figure*}
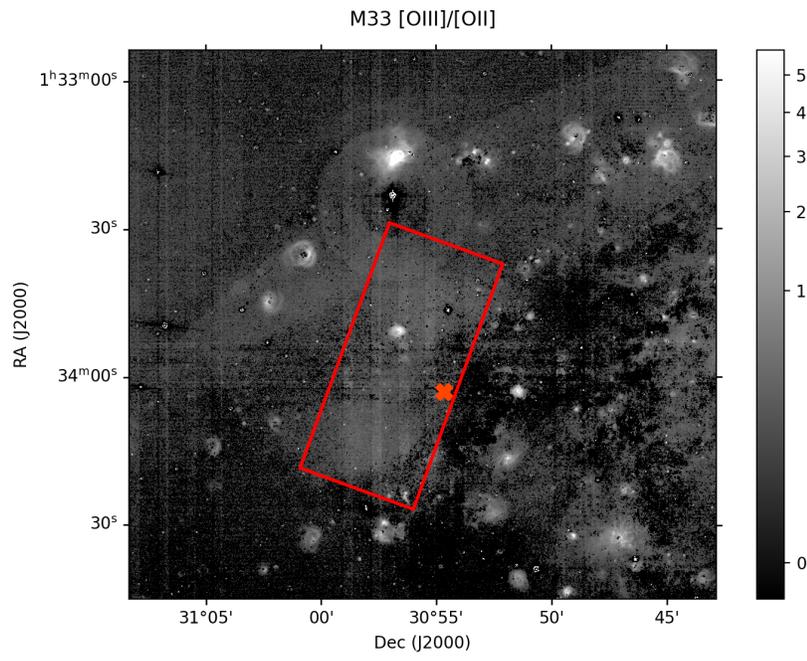

    \fig{o3_o2_power.png}{0.7\textwidth}{}
    \caption{Unsmoothed \oiii/\oii\  ratio map. The kpc-sized patch is marked. The location of unusually bright GMCs reported by \citet{Bigiel2010ApJ...725.1159B} is marked with a red X.}
    \label{fig:hii-patch}
\end{figure*}

In general, the \oiii\wavelength{5007}/\oii\wavelength{3727}\space ratio in the diffuse ISM is low, $< 0.4$, and largely invariant. However, in a region bounded approximately by R.A. $\sim$ 1h 33m 40s to 1h 34m 20s and decl. $\sim$ 30\degr 53\arcmin\ to 30\degr 59\arcmin, we observe a large-scale, bi-lobed patch of diffuse emission where the \oiii\wavelength{5007}/\oii\wavelength{3727}\space ratio is $\sim 0.6$, significantly greater than the background (Figure \ref{fig:hii-patch}). This patch corresponds to an area $\sim 9\arcmin\times3\arcmin.5$, implying a structure with dimensions on the order of $\sim$1 kpc for the M33 distance of 840 kpc \citep{Freedman1991ApJ...372..455F}. The integrated \oiii\wavelength{5007} luminosity of the diffuse patch, excluding other \hii\space regions in the vicinity, is $\sim(4.9\pm 1.5)\times10^{38}$ ergs/s. Such a large scale region of elevated ionization is unusual and difficult to explain.

There is an optically thin \hii\space region, BCLMP 637, at the centre of this patch of excited gas. BCLMP 637 is one of the most highly ionized objects in our sample, with an \oiii\wavelength{5007}/\oii\wavelength{3727}\space ratio of $\sim4$. Its ionization is apparently dominated by [NM2011] J013350.71+305636.7 \citep{NM2011ApJ...733..123N}, a WN3 star. We evaluate whether this WN3 star is responsible for photoionizing the large, excited patch in what follows.

From aperture photometry, we find that the observed \oiii\wavelength{5007} luminosity of BCLMP 637 is $\sim(1.38\pm 0.08)\times10^{38}$ erg/s, about $4\times$ less than that of the large patch.
The \Halpha\ luminosity of BCLMP 637 is $(6.75\pm 0.62) \times10^{37}\ \rm erg\ s^{-1}$, corresponding to $Q(\rm H_0) \sim 5\times 10^{49}\ \rm s^{-1}$. We compare this to the PoWR WNE stellar models at LMC metallicity by \citet{Todt2015A&A...579A..75T} and find that this value is roughly an order of magnitude higher than that predicted by the brightest models. In particular, the LMC WNE PoWR models 10-17, 09-16, and 13-21 predict $M_V$ in the range --4.8 to --5.0 for an early type WR star, which agrees with $M_V$ of --4.9 for [NM2011] J013350.71+305636.7 \citep{NM2011ApJ...733..123N}. The ionizing photon flux predicted by these models is in the range $1.1-1.3\times10^{49}$ photons/s, which is much less than the $Q(\rm H_0)$ required to ionize even BCLMP 637.  Thus, additional ionizing OB stars are likely present in the nebula.  However, there is no further evidence suggesting an unusual stellar population in this object that can be responsible for also ionizing the large, extended patch of elevated ionization.

Thus we also consider candidate high-mass X-ray binaries (HMXBs) within the patch from the X-ray survey of M33 by \citet{Pietsch2004A&A...426...11P}.
We examine two sources: [PMH2004] 192 and [PMH2004] 229.  Extrapolating the reported X-ray fluxes in the 0.2--4.5 keV band from \citet{Pietsch2004A&A...426...11P} with typical HMXB power law indices of 1.5-2.5 results in ionizing photon emission rates $Q(H_{0})$ on the order of $10^{46}\ \rm s^{-1}$. Again, this value of $Q(H_{0})$ is orders of magnitude below the required $10^{50}\ \rm s^{-1}$ to explain the origin of the elevated excitation. 

Interestingly, \citet{Bigiel2010ApJ...725.1159B} report the existence of unusually hot and bright giant molecular clouds (GMCs) adjacent to this high ionization patch, with a lower inferred CO-to-H$_2$ conversion factor. These GMCs are located between the two lobes, slightly south of center (Figure \ref{fig:hii-patch}). This suggests that the elevated ionization is indeed real and physically associated with M33, and that the source responsible for exciting the diffuse \oiii\space is also heating up the GMCs. The identity and nature of this source remains unknown.

\section{Conclusion} \label{sec:conclusion}
To summarise, we have used narrow-band imaging of M33 to generate ratio maps in \oiii\wavelength{5007}/\sii\wavelength{6724} \space and \oiii\wavelength{5007}/\oii\wavelength{3727} \space to explore the limits of ionization-parameter mapping as a probe of \hii\space region optical depth to ionizing UV radiation. We employ a revised empirical method for continuum subtraction building on methods by \citet{Keenan2017ApJ...848...12K} and \citet{Hong2014PASP..126...79H}. This method uses the pixel histogram distribution for diffuse emission after filtering out bright, resolved emission, and exploits the dispersion around the mode. 

We show that, due to the metallicity dependence of the \oiii\wavelength{5007} emission line, the \oiii\wavelength{5007}/\sii\wavelength{6724} \space and \oiii\wavelength{5007}/\oii\wavelength{3727} \space ratios can only be effective as optical depth diagnostics in the low metallicity regime ($12 + \log\left(O/H\right) \lesssim 8.4$), which is roughly $\lesssim 0.5Z_{\odot}$. Most likely, we are unable to identify a number of optically thin \hii \space regions at higher metallicities due to the weakness of \oiii\wavelength{5007} emission in this regime.  Other emission lines should be used to trace higher ionization species in these conditions.

We report the presence of a peculiar large scale ($\gtrsim 1$ kpc) structure in northern M33 that is excited in \oiii\wavelength{5007} and conspicuously absent in other bands. The known WR star and HMXBs in the vicinity of this patch cannot provide the required radiation to account for its ionization. Further observations are needed to understand its origin.

\acknowledgments

We thank Rogier Windhorst for use of the continuum filters and Anne Jaskot for assistance with the observing run.
\added{We are also grateful to the anonymous referee for helpful comments.}
This work was supported by NSF AST-1210285, NASA HST-GO-15088, and the University of Michigan.

Data processing was performed using the open-source Python libraries AstroPy, \citep{astropy:2013,astropy:2018}, Photutils \citep{photutilsBradley_2019_2533376}, NumPy \citep{numpy2020NumPy-Array} and SciPy \citep{scipy2020SciPy-NMeth}. Animations of the continuum subtraction process (Figures \ref{fig:video}, \ref{fig:haro-11-histogram}, \ref{fig:eso338-histogram} and \ref{fig:mrk71-histogram}) were prepared by generating individual frames in Matplotlib \citep{matplotlibHunter:2007} and then combined using FFmpeg \citep{ffmpegTomar2006}. Static and interactive versions of Figures \ref{fig:metal-chart} and \ref{fig:thinfraction} were created using the data visualisation libraries Altair \citep{altairVanderPlas2018} and Vega \citep{vegaSatyanarayan2017}.

\vspace{5mm}
\facilities{Mayall(MOSAIC-1), HST(STIS)}

\clearpage
\appendix 

\section{Application to \ciii\ imaging of starburst galaxies}\label{sec:other-applications}

In Section~\ref{sec:IPM}, we demonstrate the continuum subtraction method based on the modal bin fraction for the M33 data. Given the proximity of M33,  a Local Group galaxy,
individual stars are resolved, as well as a strong color gradient in the diffuse stellar background.
Applying the method to more distant galaxies with more diffuse starlight and more uniform color can provide a cleaner subtraction.  Here we demonstrate the method for continuum subtraction of faint \ciii$\lambda1909$ imaging of three starburst galaxies, ESO 338-IG004, Haro 11 and Mrk 71, using observations from the Hubble Space Telescope (HST; program GO-15088, PI Micheva). The imaging was obtained with STIS, using the F25CIII filter for the line image; and F25QTZ (Haro 11 and ESO 338) and F25CN182 (Mrk 71) for the continuum.

\begin{figure}
    \plotone{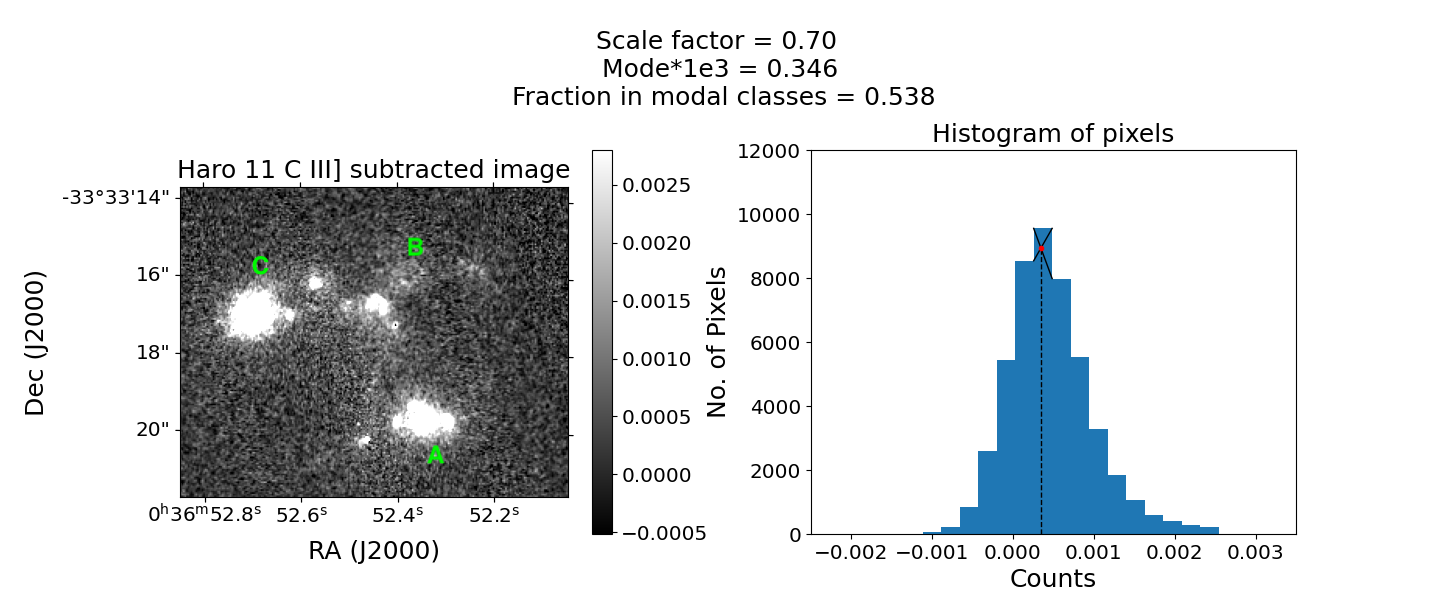}
    \begin{interactive}{animation}{haro11_CIII_movie.mp4}
        \plotone{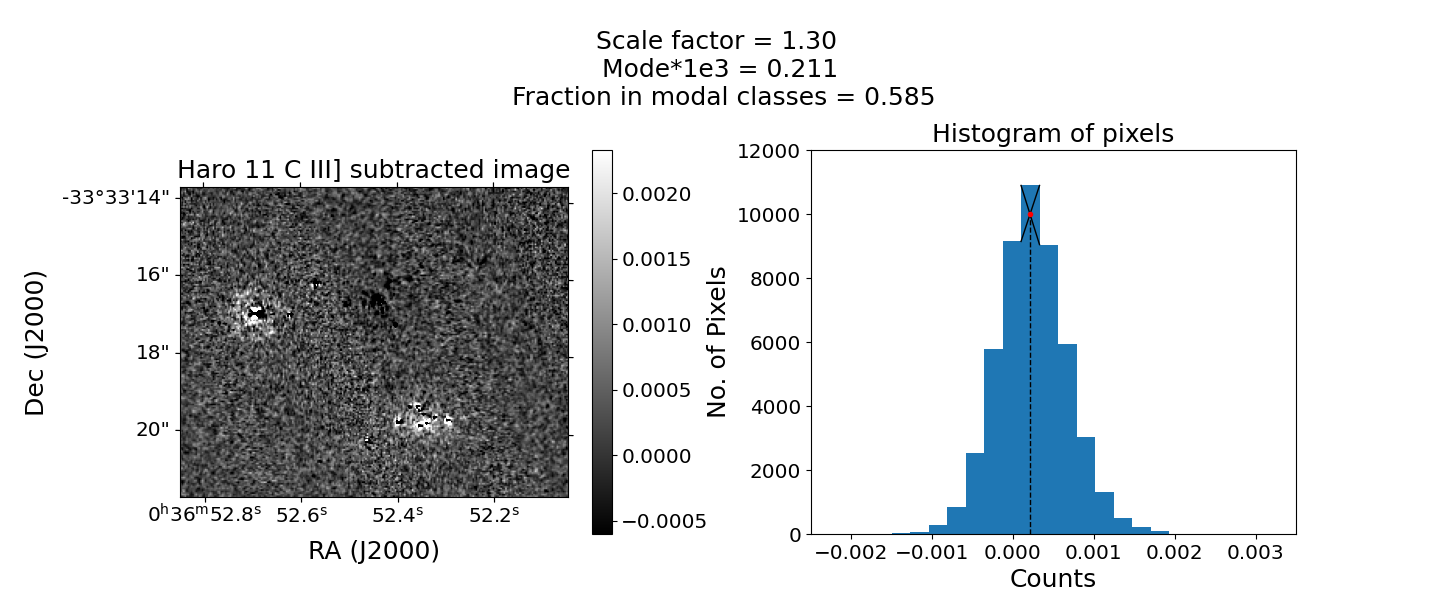}
    \end{interactive}
    \plotone{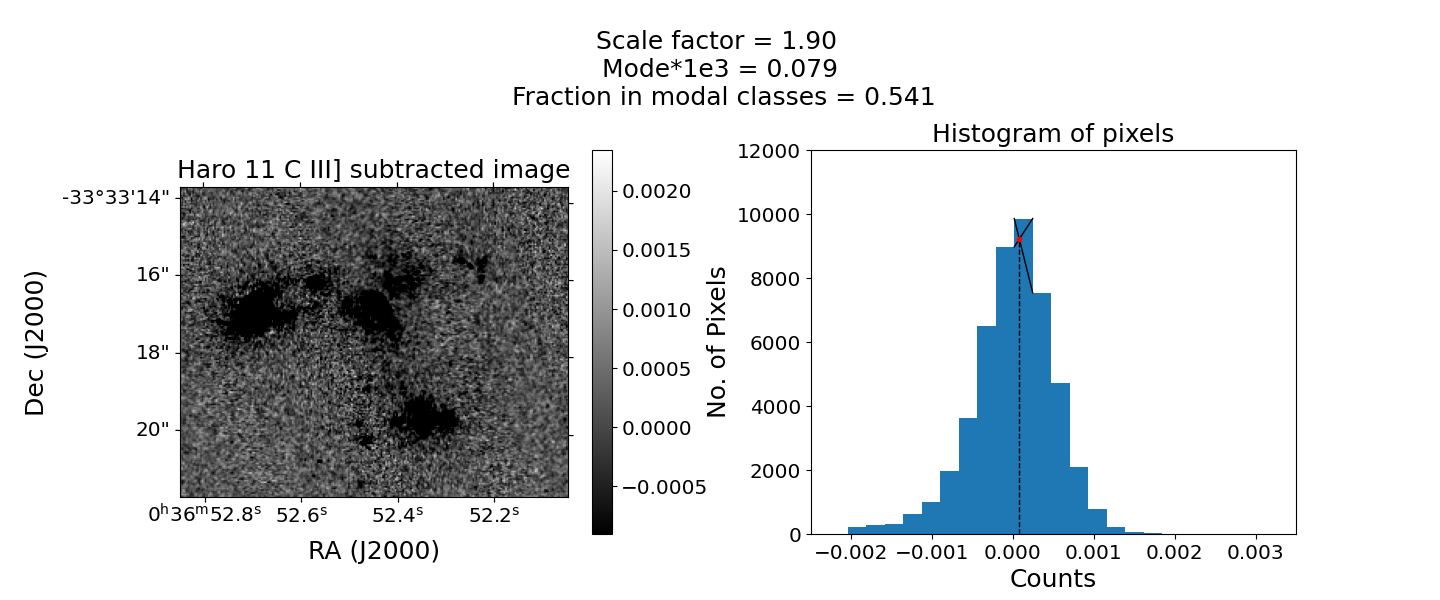}
    \caption{ Histograms showing undersubtracted, optimally subtracted, and oversubtracted images of Haro 11. Knots A, B and C are labelled in the top panel. These illustrate the transition in skewness and the tails of the histograms. An animated version of the continuum subtraction process, showing a step-by-step increase in the scale factor from 0 to 2, is available in the online version of this paper. The video duration is 20 seconds.}
\label{fig:haro-11-histogram}
\end{figure}

\begin{figure}
    \plotone{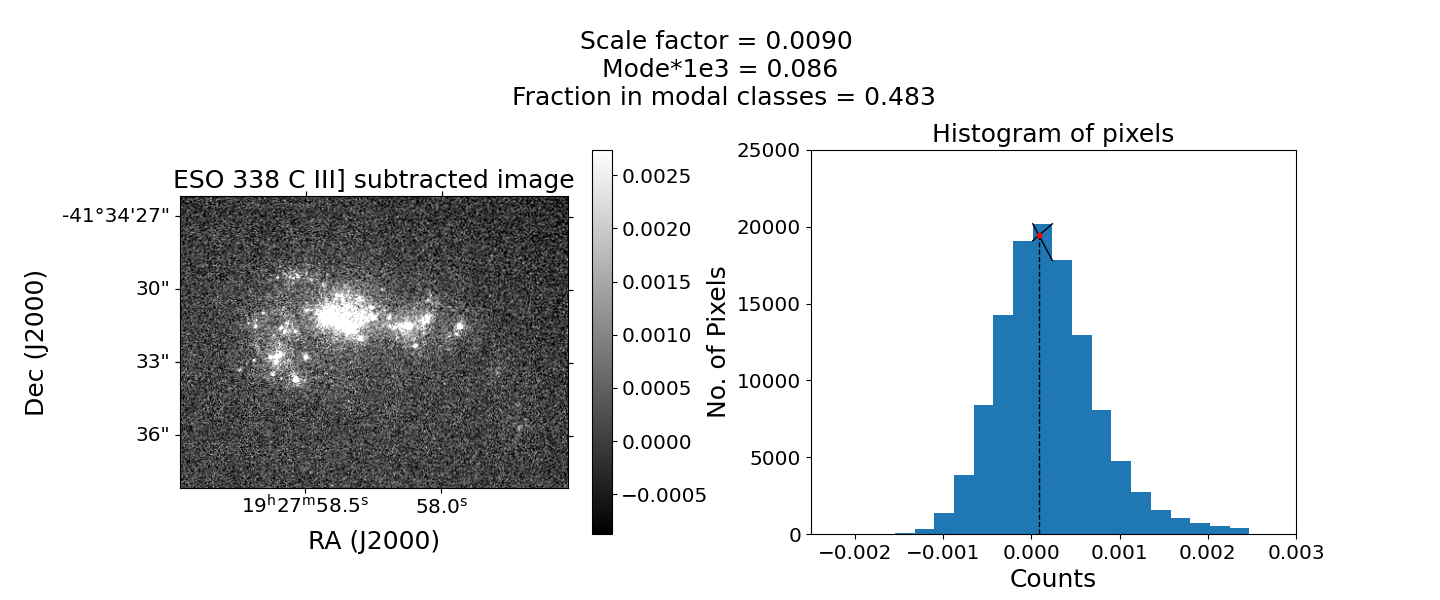}
    \begin{interactive}{animation}{eso338_CIII_movie.mp4}
        \plotone{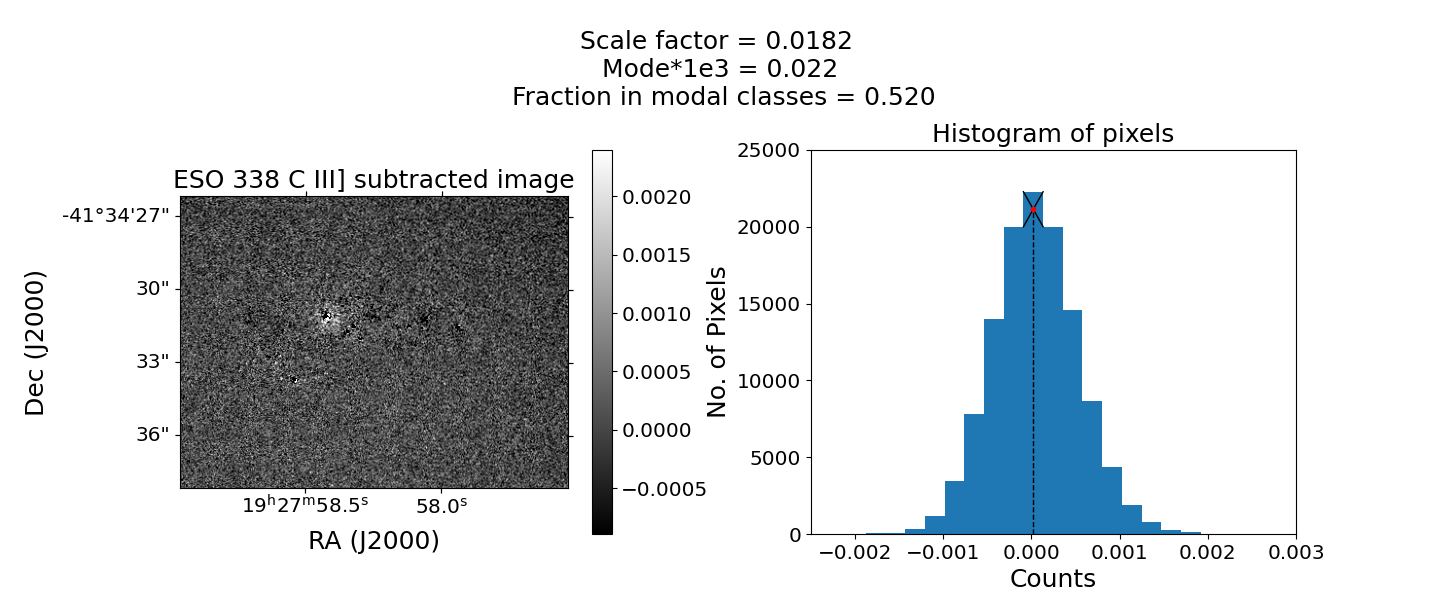}
    \end{interactive}
    \plotone{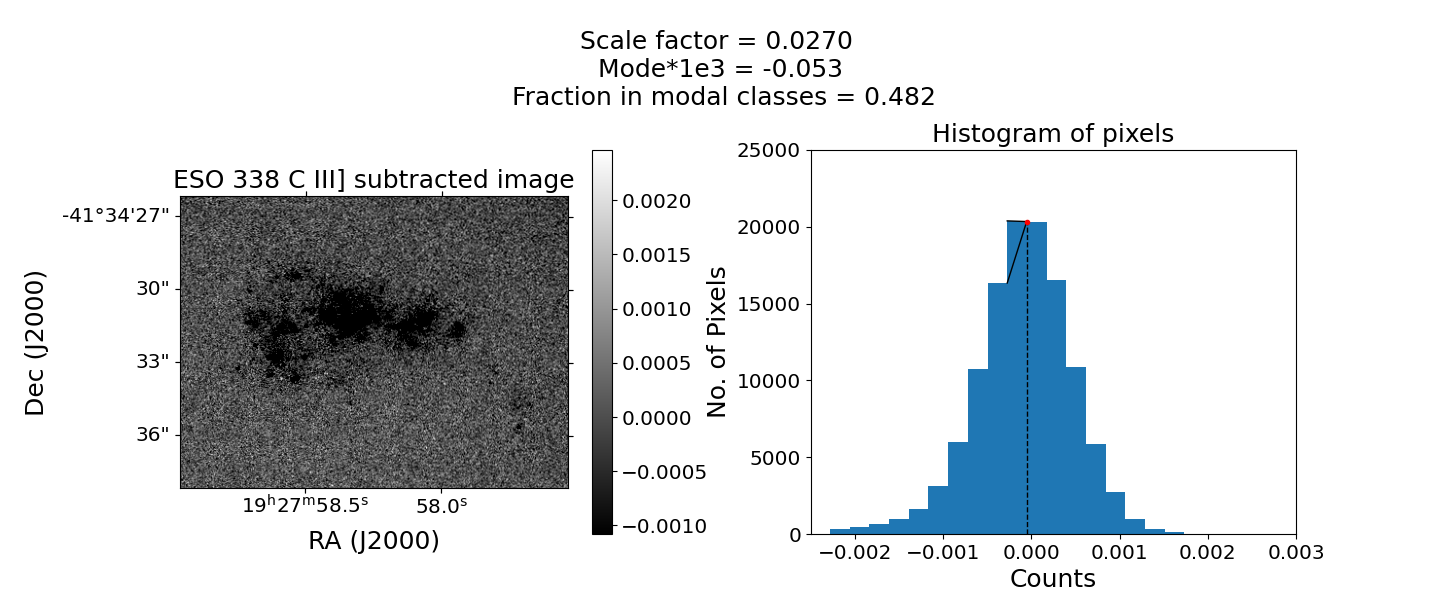}
    \caption{ Histograms showing undersubtracted, optimally subtracted, and oversubtracted images of ESO 338. An animated version of the continuum subtraction process, showing a step-by-step increase in the scale factor from 0 to 0.03, is available in the online version of this paper. The video duration is 15 seconds.}
\label{fig:eso338-histogram}
\end{figure}

\begin{figure}
    \plotone{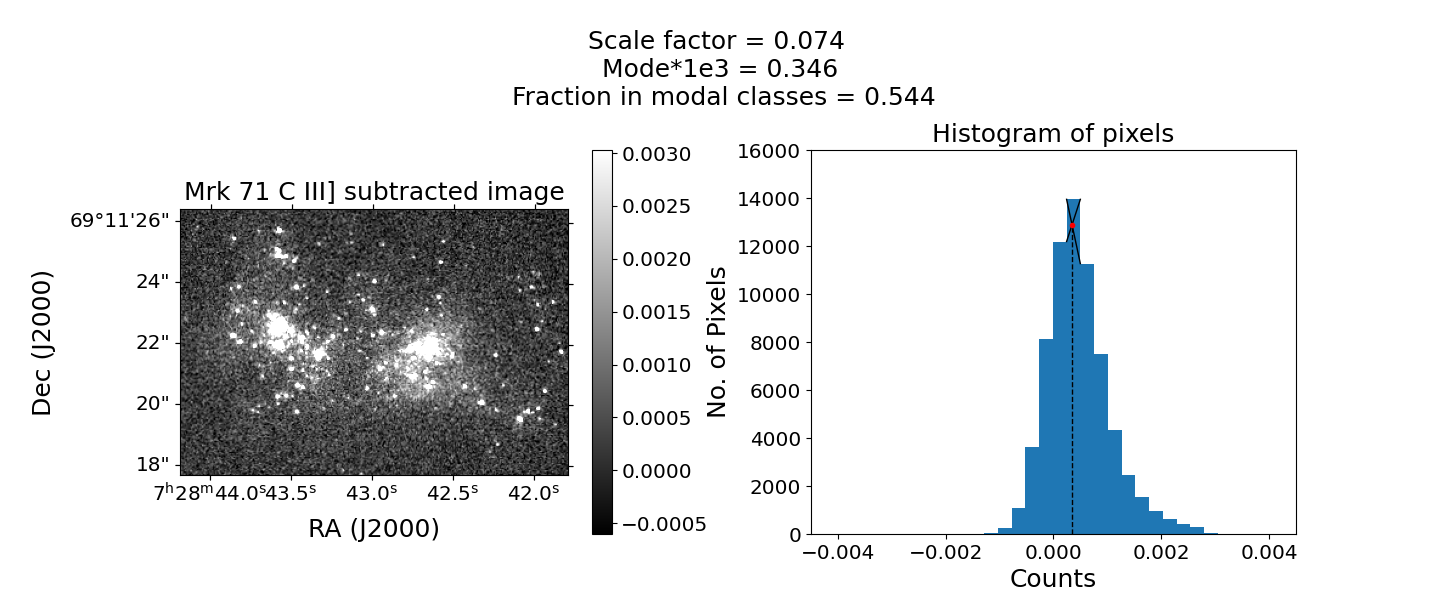}
    \begin{interactive}{animation}{mrk71_CIII_movie.mp4}
        \plotone{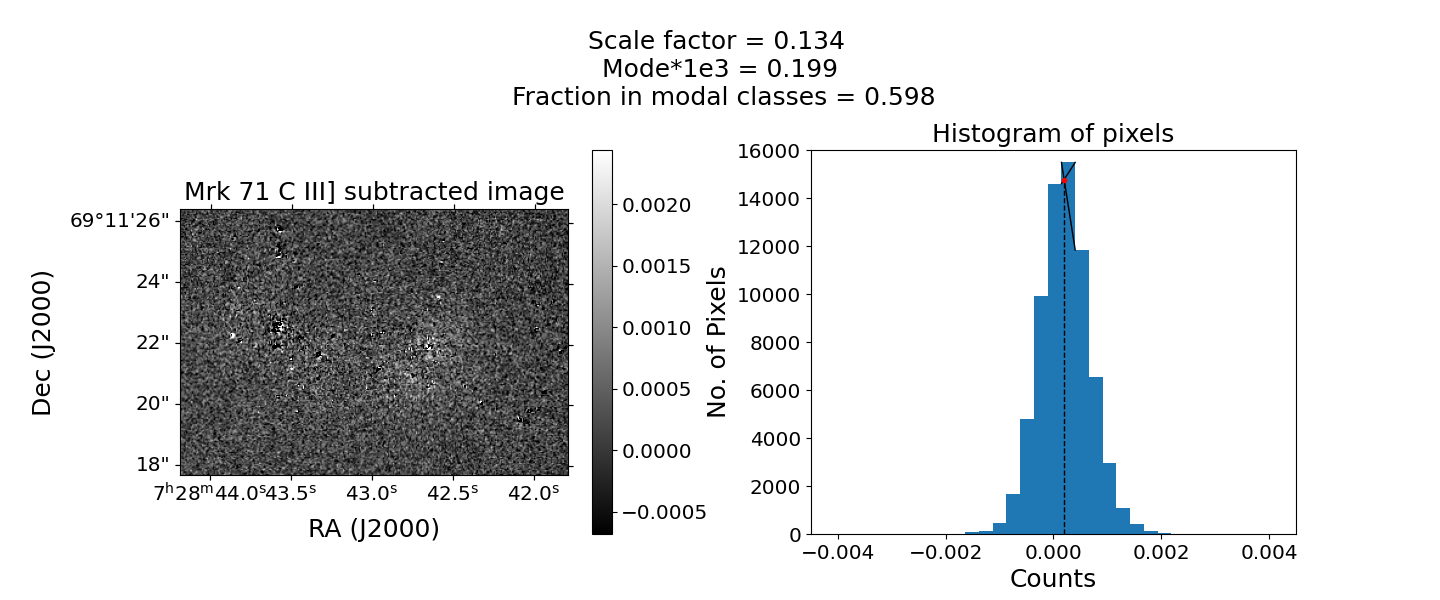}
    \end{interactive}
    \plotone{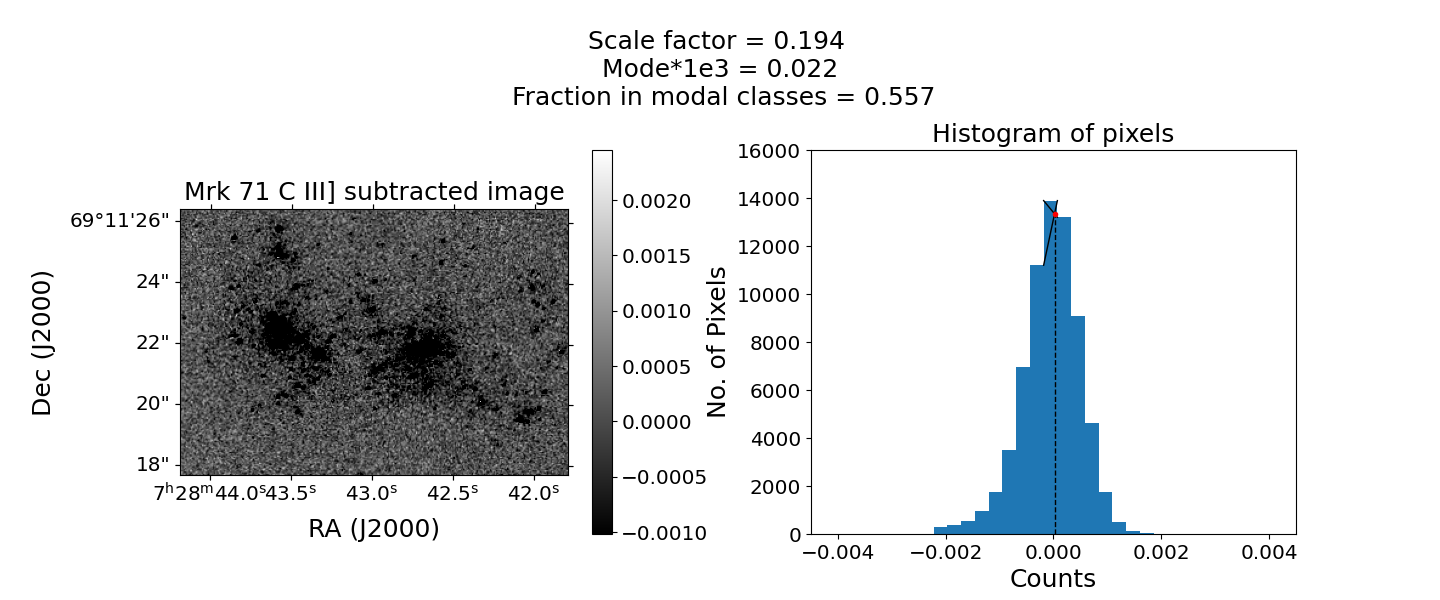}
    \caption{ Histograms showing undersubtracted, optimally subtracted, and oversubtracted images of Mrk 71. An animated version of the continuum subtraction process, showing a step-by-step increase in the scale factor from 0 to 0.3, is available in the online version of this paper. The video duration is 15 seconds.}
\label{fig:mrk71-histogram}
\end{figure}

Due to the near absence of sky emission in these HST images, the variance of background pixels is very low, and we therefore use a $4 \sigma$ clip, to isolate the diffuse emission instead of $3 \sigma$ as used for M33. As noted earlier, if sky pixels dominate the pixel population, then our method tends to produce undersubtracted images relative to diffuse background starlight. We therefore crop the region of interest to exclude empty sky regions. This amplifies the small changes to the modal bin fraction from the diffuse starlight and makes the transition from undersubtraction to oversubtraction easier to discern. Figures \ref{fig:haro-11-histogram},  \ref{fig:eso338-histogram}, and \ref{fig:mrk71-histogram} present the image subtraction at different scale factors and the pixel histograms of the continuum-subtracted images. Figures~\ref{fig:haro-11-fraction}, \ref{fig:eso338-fraction}, and \ref{fig:mrk71-fraction} show the fraction of pixels in the modal bins as function of scale factor for these galaxies, respectively.
\added{These figures also demonstrate the differences in the diagnostics between our method and those of \citet{Keenan2017ApJ...848...12K} and \citet{Hong2014PASP..126...79H}.  We see that our mode-based method provides a simpler diagnostic with greater precision than either of the other two methods.  The three methods do agree within the errors for the quantitative example in Figure~\ref{fig:haro-11-histogram}.
}
\explain{Addressing comment 1 of the referee.}

\begin{figure*}
\gridline{\fig{haro11_Mode.png}{0.3\textwidth}{(a)}
          \fig{haro11_Skew.png}{0.3\textwidth}{(b)}
          \fig{haro11_Fraction_in_modal_bins.png}{0.3\textwidth}{(c)}
          }
\caption{Haro 11 continuum subtraction: \added{Comparison of modal bin fraction with diagnostics of \citet{Keenan2017ApJ...848...12K} and \citet{Hong2014PASP..126...79H}. In panel (a), we see that there is a slope transition at scale factor $\sim0.8$ and larger oscillations for scale factor $>1.5$. The skewness metric by \citet{Hong2014PASP..126...79H} in panel (b) narrows down the interval to (1.05, 1.3), but the graph is almost a straight line. Our proposed metric in panel (c) gives an optimal scale factor in the interval (1.2, 1.4), with a simple criterion corresponding to the peak of the relation.  See also Figures~\ref{fig:eso338-fraction} and \ref{fig:mrk71-fraction}.
}}
\label{fig:haro-11-fraction}
\end{figure*}


\begin{figure*}
\gridline{\fig{eso338_Mode.png}{0.3\textwidth}{(a)}
          \fig{eso338_Skew.png}{0.3\textwidth}{(b)}
          \fig{eso338_Fraction_in_modal_bins.png}{0.3\textwidth}{(c)}
          }
\caption{Same as Figure \ref{fig:haro-11-fraction} for ESO 338.}
\label{fig:eso338-fraction}
\end{figure*}


\begin{figure*}
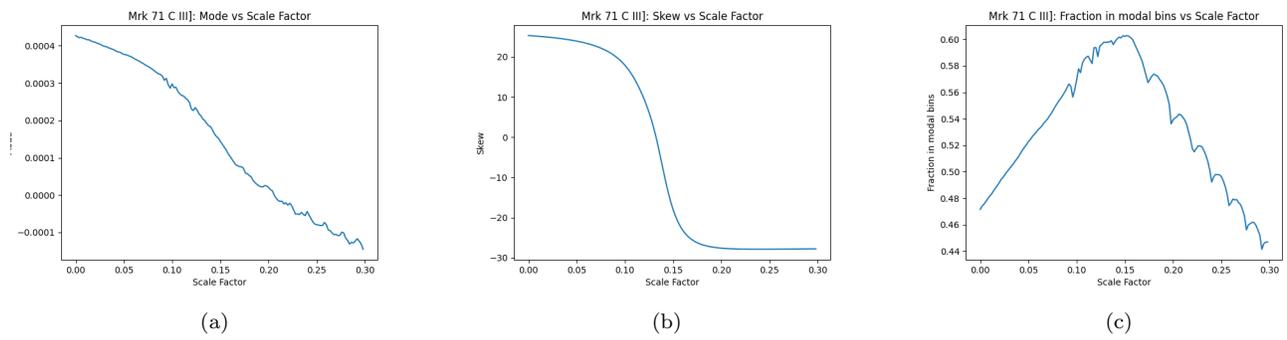

\gridline{\fig{mrk71_Mode.png}{0.3\textwidth}{(a)}
          \fig{mrk71_Skew.png}{0.3\textwidth}{(b)}
          \fig{mrk71_Fraction_in_modal_bins.png}{0.3\textwidth}{(c)}
          }
\caption{Same as Figure~\ref{fig:haro-11-fraction} for Mrk 71. }
\label{fig:mrk71-fraction}
\end{figure*}


In Figures \ref{fig:haro-11-histogram}, \ref{fig:eso338-histogram} and \ref{fig:mrk71-histogram}, we can see how the central peak shows hardly any change in shape. This is a consequence of the almost zero sky value in HST images. The tails of the histogram are well defined in the top and bottom panels, but contain less than 10\% of the total pixels. Therefore, the aperture size must be carefully chosen to avoid large swaths of empty sky. The 
pixels that hold the signal for the transition from undersubtraction to oversubtraction would otherwise be drowned out.

For Haro 11 (Figure \ref{fig:haro-11-histogram}), \citet{Keenan2017ApJ...848...12K} note that the scale factors for the three, main starburst knots differ in the continuum subtraction of narrow-band emission-line imaging, due to different-age stellar populations, so performing a smaller crop on the individual regions would be necessary for optimal continuum subtraction. 
Figure \ref{fig:haro-11-fraction}\added{(c)} shows that the algorithm predicts an overall scale factor for the entire galaxy in the interval (1.2, 1.4). The exact scale factor for each knot can be seen in the interactive version of Figure~\ref{fig:haro-11-histogram}. It is 1.38 for Knot A, and 1.26 for Knot C, both of which show residual emission at the respective scale factors whereas Knot B subtracts out completely at the predicted scale factor of 1.28. 
In Knot C, the central star cluster has bright \ciii\ line emission. Since the line and off-line images are not PSF-equalised and the MAMA detector PSF is known to have large wings, what appears as a circular region of diffuse emission  around Knot~C is 
\deleted{most likely}
an artifact of the subtraction process. \added{The radius of the observed $0\arcsec.75$ circular halo matches the expected extent of the PSF wings in the narrow band filter, given the measured S/N in the image (Figure \ref{fig:haro-11-histogram}(b)). The emission in Knot A is more likely to be real, since it} 
\explain{Addressing minor comment 2 by referee.}
has fainter clusters.  This interpretation of our results for Haro 11 is consistent with the findings of \citet{Micheva2020ApJ...903..123M}, who carried out a careful, spatially resolved analysis of the \ciii\ line vs continuum emission from this dataset.

For ESO 338 (Figure \ref{fig:eso338-histogram}), the analysis yields similar results. The peak due to background and sky pixels is broad and shows little variation with scale factor. The line and continuum pixels produce a noticeable transition in the tails of the histogram. Figure \ref{fig:eso338-fraction}\added{(c)} shows that the modal bin fraction changes only by 0.06, but the peak is well defined. The metric chosen is sensitive to small variations and can determine the optimum scale factor correctly.  The apparent diffuse emission near the strong point source in the central star burst region again may be due to the PSF wings.

Mrk 71 (Figure \ref{fig:mrk71-histogram}) contains two super-star clusters, referred to as Knot A (west) and Knot B (east) \citep[e.g.,][]{Micheva2017ApJ...845..165M}. The results of our continuum subtraction reveal that Knot A shows some faint, diffuse \ciii\space emission that is extended and therefore real. Knot B shows at least one unresolved point source, which is likely stellar \ciii\space emission from a known WC star \citep{Drissen2000AJ....119..688D}. No diffuse emission appears to be present in Knot B. Choosing the aperture so that the faint residual emission pixels are not outnumbered by background pixels is key. In Figure \ref{fig:mrk71-fraction}\added{(c)}, we see for the chosen aperture size that the algorithm detects the correct scaling without producing oversubtraction.

The \ciii\ emission in Haro 11-A and Mrk 71-A are both associated with the highest ionization-parameter regions in these two galaxies.  \citet{Gray2019ApJ...887..161G} suggest that enhanced \ciii\ may be associated with the suppression of adiabatic mechanical feedback by strongly cooling outflows. We note that Mrk 71-A has been suggested to show evidence of such suppressed superwinds \citep{Oey2017ApJ...849L...1O}.

Thus, the new continuum subtraction method based on the modal bin fraction can be applied to a variety of situations as long as there is a continuous distribution of pixels and a healthy mix of emission, continuum and sky pixels. 
\added{Empirical methods work best when there are $\gtrsim 10^4$ pixels where the majority are dominated by stellar continuum and not contaminated by line emission.}
\explain{Addressing comment 1.1 of referee.}

\bibliography{main}{}
\bibliographystyle{aasjournal}
\listofchanges
\end{document}